# The slip deficit on the North Anatolian Fault (Turkey) in the Marmara Sea: Insights from paleoseismicity, seismicity and geodetic data.

M. Meghraoui[1,2*], R. Toussaint[1,2,3] and M. E. Aksoy[4]

1 - Université de Strasbourg, CNRS, ITES, UMR 7063, 67084 Strasbourg, France
2 - Université de Strasbourg, CNRS, EOST, UAR 830, 67084 Strasbourg, France
3 - SFF PoreLab, The Njord Centre, Department of Physics, University of Oslo, Oslo, Norway
4 - Department of Geological Engineering, Muğla Sıtkı Koçman University, Muğla, Turkey
*m.meghraoui@unistra.fr

**Abstract.**

The North Anatolian Fault experienced large earthquakes with 250-400 years recurrence time. In the Marmara Sea region, the 1999 (Mw = 7.4) and the 1912 (Mw = 7.4) earthquake ruptures bound the Central Marmara Sea fault segment. Using historical-instrumental seismicity catalogue and paleoseismic results ($\simeq$ 2000-year database), the mapped fault segments, fault kinematic and GPS data, we compute the paleoseismic-seismic moment rate and geodetic moment rate. A clear discrepancy appears between the moment rates and implies a significant delay in the seismic slip along the fault in the Marmara Sea. The rich database allows us to identify and model the size of the seismic gap and related fault segment and estimate the moment rate deficit. Our modeling suggest that the locked Central Marmara Sea fault segment even including a creeping section bears a moment rate deficit $\dot{M}_d$ = 6.4 x $10^{17}$ N.m./yr that corresponds to Mw $\simeq$ 7.4 for a future earthquake with an average $\simeq$ 3.25 m coseismic slip. Taking into account the uncertainty in the strain accumulation along the 130-km-long Central fault segment, our estimate of the seismic slip deficit being $\simeq$ 10 mm/yr implies that the size of the future earthquake ranges between Mw = 7.4 and 7.5.

# 1 – Introduction

The North Anatolian Fault (NAF) is a major east-west trending, continental right-lateral strike-slip fault that limits Anatolia from Eurasia and extends for 1600 km from the Karliova triple junction to the Sea of Marmara region (Barka and Kadinsky-Cade, 1988; Sengör et al., 2014). Considered as a major transtensional step-over, the Sea of Marmara experienced in the past, several large earthquakes with Mw > 7.0 (Ambraseys, 2002; Armijo et al., 2005). The most recent large seismic events that occurred West and East of Marmara Sea are the 1912 Ganos (Mw = 7.4) and 1999 Izmit (Mw = 7.4), respectively, both associated with surface ruptures that extended into the sea (Barka, 1999; Barka et al., 2002; Altunel et al., 2004; Aksoy et al., 2010, Uçarkuş et al., 2011). Although the coseismic surface ruptures helped in the identification of individual fault segments, a major issue is the fault segment dimensions, the size of seismic gap and related faulting behaviour in the Marmara Sea.

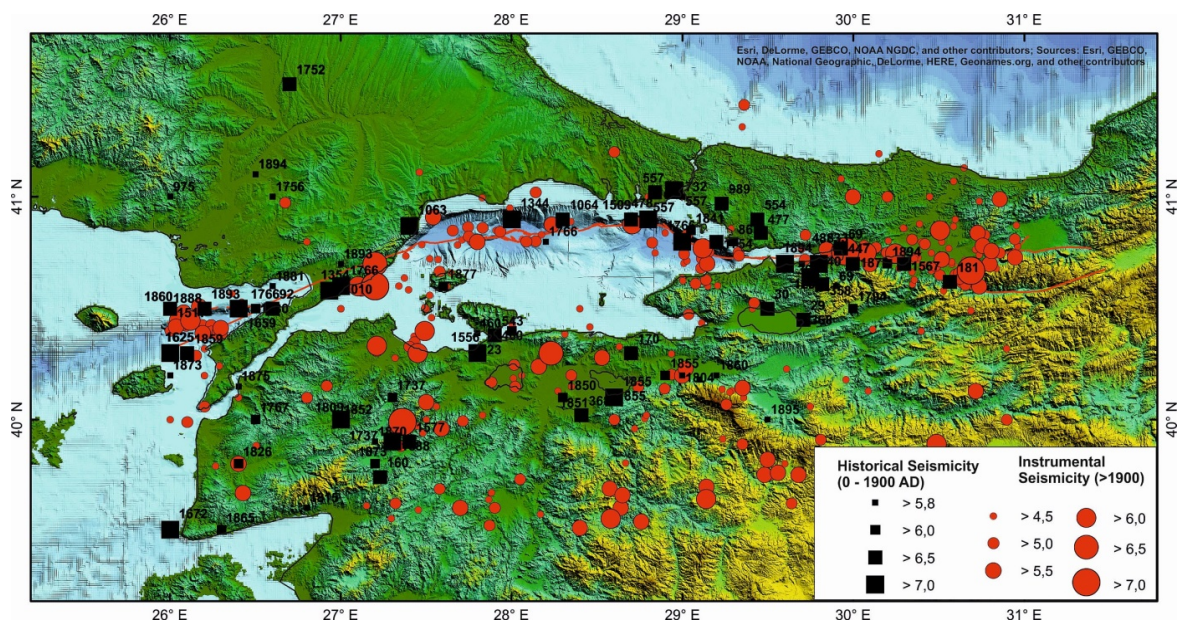

*Figure 1. Instrumental (circle) and historical (box) seismicity, and North Anatolian Fault zone (red line) in the Marmara Sea region. The instrumental seismicity data (1900 - 2014) are from Kandilli Observatory and Earthquake Research Institut and the International Seismological Center , and historical earthquakes are from Ambraseys (2002), Atakan and Sorensen (2002).*

Numerous post-1999 studies refer to the 100-150 km-long seismic gap in the Marmara Sea inferred from the instrumental and historical earthquake catalogue. Previously, the modeling of fault interaction using the Coulomb failure function provided a clear loading stress in the sea (Stein et al., 1997). Since the 1999 Izmit earthquake, the Marmara Sea and surrounding areas have been the site of extensive investigations in fault zone mapping, paleoseismology, Global Positioning System (GPS), prospection geophysics, sea floor geodesy and topography, coring and seiche investigations, gas emanation along fault ruptures (Le Pichon et al., 2003; Reilinger et al., 2006; Laigle et al., 2008; Flerit et al. 2014; Geli et al., 2008; Meghraoui et al., 2012).

The Marmara Sea seismic gap being the major issue, the data analysis and modeling were mainly dedicated to the better knowledge of the rupture zone and related seismic hazard and risk assessment of the Istanbul major city and region (Hubert-Ferrari et al., 2000; Meade et al., 2002; Hergert and Heidbach, 2010). However, in spite of several studies the location of tectonic loading on the NAF is still in debate. A major issue is if the Central segment releases its stress by creeping mode, totally or partially (Ergintav et al., 2014; Kido et al., 2017; Klein et al., 2017; Aslan et al. 2019) or is it fully locked (Sakic et al., 2016). East of Marmara Sea, along the 1944 rupture segment (Mw = 7.4) of the NAF, the Ismetpasa example shows that locked-fault and creeping section are not incompatible (Cetin et al., 2014), and suggests that the occurrence of large earthquakes in Istanbul area, remains possible. The seismic gap was primarily identified according to the seismicity catalogue but often based on the damage distribution of historical earthquakes (Ambraseys, 2002). Beside the 1999 and 1912 large events, major earthquakes that affected the Marmara Sea region are the 1344, 1354, 1509, 1719, 1766 a and b and 1894 (Fig. 1, Table 1). However, eastern and western limits of the seismic gap and related Central Marmara fault segment are not clearly identified.

In this paper, the use of a completed seismicity catalogue (Atakan and Sorensen, 2002; Ambraseys, 2009; Karabulut et al., 2011), detailed fault mapping (Le Pichon et al., 2003), paleoseismic record (Klinger et al., 2003; Rockwell et al., 2009; Meghraoui et al., 2012; Dikbas et al., 2018), and GPS data help us to constrain the extent of the seismogenic central Marmara fault segment. The seismic gap being intimately related to the seismic slip deficit and taking into account the adopted recurrence time of large events, we use the difference between the geodetic moment rate and the seismic-paleoseismic moment rate in order to quantify and model the location and amount of seismic slip deficit in the Marmara Sea. Our results imply the physical identification of the seismic gap and related slip deficit that we discuss with the implication for the seismic hazard assessment.

| Year | Long. (°) | Lat. (°) | Mw | Mo (N.m) |
|---|---|---|---|---|
| 478  | 28.8  | 40.9  | 7.1  | $5.69*10^{19}$ |
| 554  | 29.44 | 40.9  | 6.5  | $7.16*10^{18}$ |
| 989  | 29.25 | 40.9  | 6.5  | $7.16*10^{18}$ |
| 1063 | 27.4  | 40.87 | 7.4  | $1.60*10^{20}$ |
| 1344 | 28.1  | 40.9  | 7.1  | $5.69*10^{19}$ |
| 1354 | 26.93 | 40.58 | 7.4  | $1.60*10^{20}$ |
| 1509 | 28.7  | 40.9  | 6.75 | $1.70*10^{19}$ |
| 1719 | 29.8  | 40.7  | 7.4  | $1.60*10^{20}$ |
| 1766 | 27    | 40.6  | 7.3  | $1.14*10^{20}$ |
| 1766 | 29    | 40.8  | 7.4  | $1.60*10^{20}$ |
| 1894 | 29.6  | 40.7  | 6.8  | $9.55*10^{19}$ |
| 1912 | 27.2  | 40.6  | 7.4  | $1.60*10^{20}$ |
| 1967 | 30.69 | 40.67 | 7.25 | $9.55*10^{19}$ |
| 1999 | 29.96 | 40.76 | 7.3  | $1.14*10^{20}$ |

Table 1: Major seismic events (Mw > 6.5) in the Marmara area (Ambraseys, 2002, 2009).

## 2 - A multidisciplinary set of data

### 2.1 Fault segmentation and seismicity

The northern branch of the NAF in the Marmara Sea accommodates most of the active deformation in the region and generates large earthquakes (Mw > 7.0) with coseismic slip reaching ≃ 5 m (Ambraseys and Jackson, 2000; Reilinger et al., 2006). The most recent earthquake with Mw = 7.4 occurred in 1999, breaking in Izmit region the eastern fault segment of the Marmara Sea region (Barka et al., 2002; Cakir et al., 2003). The background seismicity and 1999 aftershock distribution at depth allow us to estimate a ≃ 15 km-thick seismogenic layer (Karabulut et al., 2011; Schmittbuhl et al., 2016). The instrumental seismicity data obtained from Kandilli Observatory and Earthquake Research Institute and the International Seismological Center, the historical earthquake intensity data collected by Ambraseys (2002), and Atakan and Sorensen (2002) coupled with paleoseismological studies (Rockwell et al., 2009; Aksoy et al., 2010; Meghraoui et al., 2012; Dikbas et al., 2018) allow to cover almost 2000 years of a completed earthquake record with Mw > 6.0 in the Marmara area. This temporal seismicity coverage spans several seismic cycles and provide the possibility to constrain the faulting behaviour and determine the seismic gap in the Marmara Sea (Cakir et al. 2014).

Previous tectonic geomorphology and paleoseismology studies show an average of ≃ 3 m of seismic slip for both 1912 and 1999 earthquakes (Altunel et al., 2000; Altunel et al., 2004; Rockwell et al. 2009; Meghraoui et al. 2012; Dikbas et al. 2018). Based on historical documents and contemporaneous photographs, surface ruptures of the 1912 earthquake were first mapped by Altunel et al. (2000). Fault offset are also documented from the highresolution sea floor bathymetry and seismic profiles (Gasperini et al. 2011; Ucarkus et al., 2011; Yakupoğlu et al. 2019). The well identified 1999 Izmit rupture was about 140 km long (Barka et al. 2002; Cakir et al. 2003), which was comparable to the 1912 event (Mw = 7.4) (Ambraseys, 2002; Aksoy et al. 2010) and where only the 45 km-long inland were observed. Submarine fault scarps and lateral displacements of ≃ 6 m in the Tekirdag basin, suggest that the 1912 rupture extends for a total of ≃ 130 km from the Central pull-apart basin in the western Marmara Sea, till the Saros

Bay in the Aegean Sea (Rockwell et al., 2009; Armijo et al., 2005; Aksoy et al., 2010; Meghraoui et al., 2012). However, an earthquake rupture length can be largely independent from fault segment dimensions and their endpoints as identified through the geometrical complexities (pull-apart basin, step-over, fault bend; Wesnousky, 2006). In the Marmara Sea, and in between the 1999 and 1912 fault segments, a 100 to 150-km-long fault is suspected to have generated an estimated magnitude 7.4 in May 1766 (Armijo et al., 2005; Meghraoui et al., 2012). Therefore, we consider that the Central fault segment dimensions and discontinuities can be variable suggesting a range of moment-magnitudes for a future earthquake. The location of major events and coseismic surface ruptures implies the existence of three main fault segments (Fig. 1, Table 1) in the Marmara Sea, with a long-term 16 - 18 mm/yr slip rate, and 250-400 years recurrence interval according to the paloseismic results (Klinger et al., 2003; Rockwell et al., 2009; Meghraoui et al., 2012, Zabci et al. 2019).

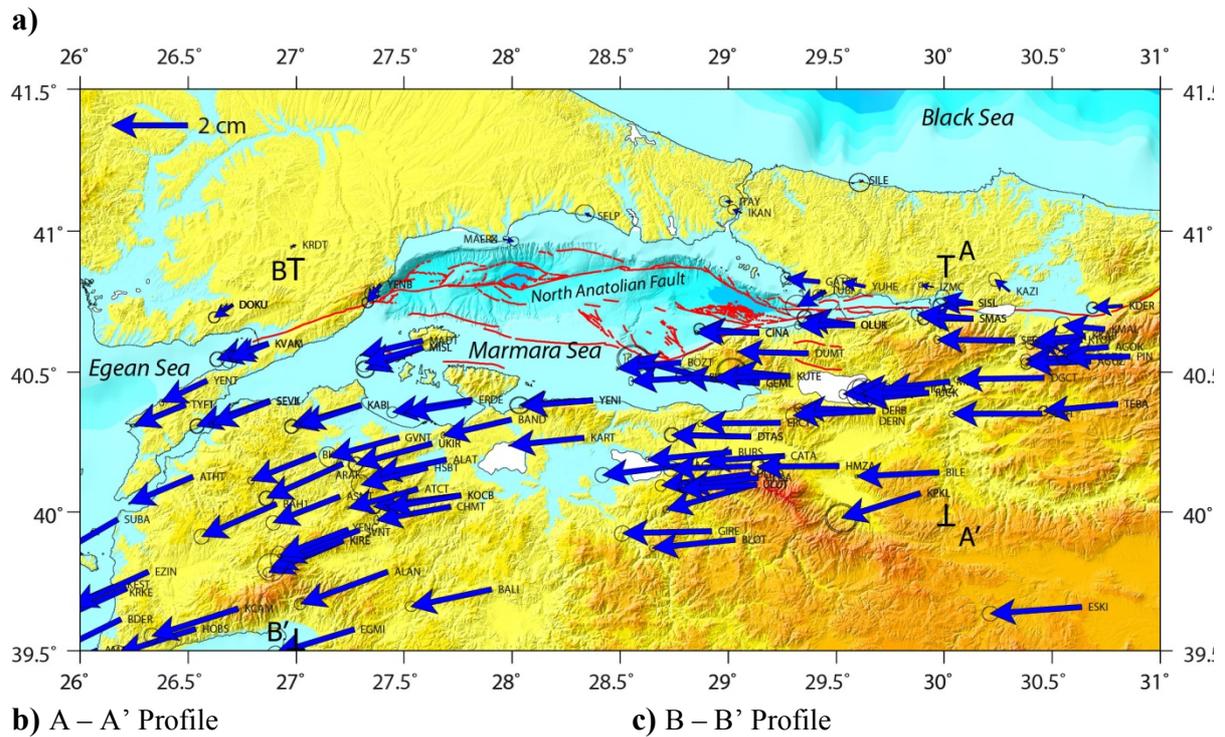

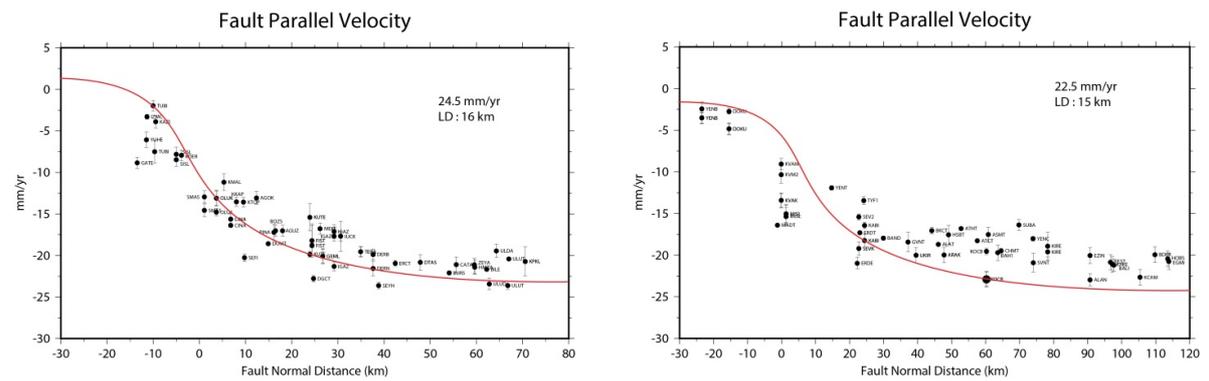

*Figure 2. **a** GPS velocities with stable Eurasia as reference from Reilinger et al. (2006), Aktug et al. (2009) and Ergintav et al. (2014); red line is the Northern Marmara Fault zone (Le Pichon et al. 2003; Armijo et al. 2005). **b** Profile A–A' of GPS velocities with a total 24.5 mm/year for a 16 km locking depth (LD). **c** Profile B–B' of GPS velocities with a total 22.5 mm/year for a 15 km locking depth (LD).*

## 2.2 Geodetic deformation field

The tectonic loading responsible of earthquake generation and slip deficit along the NAF in the Marmara Sea, is here documented with the local deformation on the fault (seismicity and paleoseismology) and compared to the regional deformation across the fault (GPS measurements). The far field active deformation and velocity field that allow the progressive stress and strain accumulation along the fault are obtained from the GPS studies (Figure 2a; Reilinger et al., 2006; Aktug et al., 2009; Ergintav et al., 2014). The increase of GPS stations during the last 20 years, with numerous measurements campaigns provide 20-25 mm/yr deformation rate in the Marmara Sea region (Reilinger et al., 2006). GPS measurements demonstrate that the NAF accommodates the westward extrusion and rotation of the Anatolian block by its right-lateral strike-slip motion with regard to the Eurasian plate (Flerit et al., 2014).

To constrain the tectonic and seismic loading along the fault, we use 197 station measurements from the three most recent studies that cover major fault segment of the Marmara Sea region (Reilinger et al., 2006; Aktug et al., 2009; Ergintav et al., 2014). Although GPS stations are not distributed homogeneously in the region (lack of stations in the sea and NW area of Marmara), geodetic measurements cover most part of the eastern 1999, the western 1912 fault segments and the southern block of the NAF (Figure 2 a). The sea itself has apparently no GPS seafloor stations, but recent acoustic extensometer measurements during 1.5 year reveal a highly locked fault zone with strain accumulation across the Central fault segment (Sakic et al., 2016; Petersen et al., 2019; Lange et al., 2019). On the other hand, Ergintav et al. (2014) use GPS velocities from extrapolated profiles in the sea compared to the elapsed time since 1766 to conclude for creeping in the central Marmara Sea fault segment. Similarly, Klein et al. (2017) use Bayesian modelling with the inversion of GPS measurements north of the Central segment and infer for creeping in the central segment. The lack of seafloor GPS measurements imposes, however, the use of southern block GPS stations, from which we obtain completed geodetic profiles with 22.5 ±0.5 mm/yr and 24.5 ±0.5 mm/yr for western and eastern profiles, respectively (see profiles A-A' and B-B' in Figures 2 b and c). We also observe that as for the Ismetpasa section of the Gerede fault segment, a creeping section and a locked rupture zone may coexist along a fault segment without precluding the possibility of a Mw > 7.0 earthquake (Cetin et al., 2014).

**3 - Seismic slip deficit evaluation in the Marmara Sea gap**

3.1 Methodology

To determine and quantify the seismic gap along the NAF, we use a similar method than that proposed by Murray and Segall (2002) along the San Andreas Fault. Here we use simultaneously the seismic (long-term, 2000 years) and the geodetic moment rate to determine the deficit moment rate for different time intervals (Ti) (Equation 1, 2).

$$\dot{M}_d = \mu \iint_s (\dot{U}_\infty - \dot{U}) ds \qquad (1)$$

where $\dot{M}_d$ is the deficit moment rate ; μ the rigidity modulus (= $3.3*10^{10}$ N.m$^2$) ; $\dot{U}_\infty$ the geodetic deformation rate, and $\dot{U}$ the coseismic slip rate on fault (Murray and Segall, 2002). Equation (1) also gives:

$$\text{Ti} = M_o / \dot{M}_d \qquad (2)$$

where Ti is the return period of large earthquakes, $M_o$ is the cumulative seismic moment over the considered time interval. The deficit moment rate $\dot{M}_d$ corresponds to the difference between the deformation rate far from the fault, due to the tectonic plate motion and the deformation along the fault due to repeated coseismic slip. The slip deficit is represented through the development of a Matlab program showing the calculations on a regular 0.3° cell. We first use seismic and paleoseismological data to determine the distribution of the seismic moment rate along the NAF (Fig. 3). We consider the seismic moment of each event and the total energy released on a simple fault plane to calculate the cumulative seismic moment per segment ($\Sigma M_0$) and infer the seismic strain velocity for ≃ 2000 year-long of the seismicity catalogue (Equation 3; Kostrov, 1974).

$$\dot{u} = \frac{1}{\mu AT} \sum_{k=1}^{n} M_0^{(k)} \qquad (3)$$

where *T* is the time interval considered. This approach allows to take into account physical rupture parameters (Aki and Richards, 1980) such as the rigidity modulus ($\mu$ = 3.3 x $10^{10}$ N.m$^2$), the coseismic displacement (*u*) and the slip area (*A* = *W L* with *W* : seismogenic width ; *L* : rupture length). Second, we observe the far field deformation along the NAF, from the 1999 to 1912 rupture segment by the calculation of the geodetic moment rate. To do so, we use Savage and Simpson (1997) method that documents the deformation in the volume across the fault trace. We take in consideration the regular grid to interpolate GPS data in less covered area and calculate the deformation for each cell.

Specifically, this is done by interpolating every grid point velocity from the GPS data and the positions of the GPS measurement points, using weights decreasing exponentially with angle of separation, and a characteristic angle of decrease corresponding to the 0.3° lattice step of the regular grid. The GPS measured velocities and their positions are shown as the vector field with the resulting interpolated vector field on a regular grid (Figs. 4a, b).

From this interpolated field $\dot{\boldsymbol{u}}_{i,j}$ on the grid points *(i,j)*, the surface strain rate tensor $\bar{\bar{\varepsilon}} = (\nabla \dot{\boldsymbol{u}} + (\nabla \dot{\boldsymbol{u}})^T)/2$ is evaluated on every grid point by the central difference method. The details of this interpolation are in the Supplementary Material. The two eigenvalues of the strain rate tensor in decreasing order are noted ($\varepsilon_1, \varepsilon_2$), and correspond to two orthogonal eigendirections. They are determined at every grid point. The resulting moment rate deficit for strike slip and dip slip components are in Figs. 5 a and b, respectively, and strain rate distribution is shown on Figs. 6a, b.

To calculate the geodetic moment rate, the method considers the diagonalized deformation tensor (Equation 4; Savage and Simpson, 1997), the double-couple mechanism of the fault, and in our case combining the strike-slip and dip-slip mechanisms (Equation 5). The NAF has a principal right-lateral strike-slip motion, but the presence of geometrical discontinuities (pull-apart, step-over) implies the existence of transtension and transpression tectonic mechanisms. The geodetic moment rate is obtained following the relation (see also Supplementary Material for more explanations on the geodetic slip partitioning):

$$\dot{M}_o = 2\mu HA \begin{bmatrix} \varepsilon 1 & 0 & 0 \\ 0 & \varepsilon 2 & 0 \\ 0 & 0 & -\Delta \end{bmatrix} \quad \text{with } \Delta = \varepsilon_1 + \varepsilon_2 \quad (4)$$

and the combined moments for the strike-slip and dip-slip mechanisms are expressed as Savage and Simpson (1997):

$$\dot{M}_{o\,(T)} = \dot{M}_{o\,(SS)} + \dot{M}_{o\,(DS)} = 2\mu HA \ \max(|\varepsilon_1|, |\varepsilon_2|, |\varepsilon_1+\varepsilon_2|) \quad (5)$$

Individually, the moments of the strike-slip and dip-slip mechanisms are expressed as follows (see supplementary information for a detailed justification built on the analysis of Savage and Simpson (1997):

For every surface cell where the geodetic moment tensor is measured, the minimum among $(|\varepsilon_1|+|\varepsilon_1+\varepsilon_2|, |\varepsilon_2|+|\varepsilon_1+\varepsilon_2|, |\varepsilon_1|+|\varepsilon_2|)$, is evaluated, and three cases arise depending on which of these three strain rates is minimum:

(a) If $|\varepsilon_1|+|\varepsilon_1+\varepsilon_2|$ is the minimum, then the geodetic scalar moment rate in strike-slip mechanism is

$$\dot{M}_{o\,(g,\,SS)} = 2\mu HA\ |\varepsilon_1| \quad (5)$$

and the geodetic scalar moment rate in dip-slip mechanism is

$$\dot{M}_{o\,(g,\,DS)} = 2\mu HA\ |\varepsilon_1+\varepsilon_2| \quad (6)$$

with a total geodetic moment rate

$$\dot{M}_{o\ (g,\ T)} = \dot{M}_{o\ (g,\ SS)} + \dot{M}_{o\ (g,\ DS)} = 2\mu HA(\ |\varepsilon_1| + |\ \varepsilon_1 + \varepsilon_2|) \qquad (7)$$

(b) If $|\varepsilon_2| + |\varepsilon_1 + \varepsilon_2|$ is the minimum, then the geodetic scalar moment rate in strike-slip mechanism is

$$\dot{M}_{o\ (g,\ SS)} = 2\mu HA\ |\varepsilon_2|, \qquad (8)$$

and the geodetic scalar moment rate in dip-slip mechanism is

$$\dot{M}_{o\ (g,\ DS)} = 2\mu HA\ |\ \varepsilon_1 + \varepsilon_2|, \qquad (9)$$

with a total geodetic moment rate

$$\dot{M}_{o\ (g,\ T)} = \dot{M}_{o\ (g,\ SS)} + \dot{M}_{o\ (g,\ DS)} = 2\mu HA(\ |\varepsilon_2| + |\ \varepsilon_1 + \varepsilon_2|) \qquad (10)$$

(c) If $|\varepsilon_2| + |\varepsilon_2|$ is the minimum: There is no geodetic scalar moment rate in strike-slip mechanism, i.e.

$$\dot{M}_{o\ (g,\ SS)} = 0 \qquad (11)$$

and the geodetic scalar moment rate in dip-slip mechanisms is

$$\dot{M}_{o\ (g,\ DS)} = 2\mu HA\ |\varepsilon_1| + |\varepsilon_2| \qquad (12)$$

with a total geodetic moment rate

$$\dot{M}_{o\ (g,\ T)} = \dot{M}_{o\ (g,\ SS)} + \dot{M}_{o\ (g,\ DS)} = 2\mu HA(\ |\varepsilon_1| + |\varepsilon_2|) \qquad (13)$$

The difference between the geodetic moment rate and the seismic moment rate ($\dot{U}_\alpha - \dot{U}$) corresponds to the deficit moment rate from which we compute the slip deficit during a seismic cycle. The coupled fault section coincides with the area of seismic slip deficit, and hence with the seismic gap, along a fault segment. It is then possible to estimate, for different time intervals (Equation 2), and from the cumulative seismic moment the associated magnitude Mw (Kanamori, 1977), according to the recurrence interval on the NAF (250 to 400 years) (Rockwell et al., 2009; Meghraoui et al., 2012; Dikbas et al., 2018).

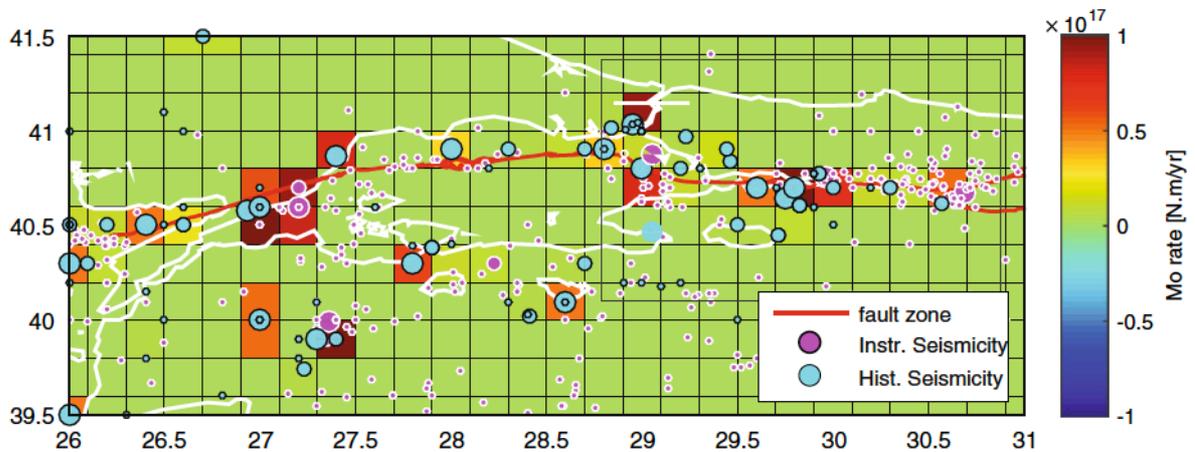

*Figure 3. Seismic moment rate obtained from ≃ 2000 year-long seismicity catalogue using Mw > 6.5 (see also Fig. 1 and Table 1). A seismic moment rate gap appears along the Central region of the North Anatolian Fault in the Marmara Sea.*

## 4 - Results

The cumulative seismic moment obtained from ≃ 2000 years-long seismicity catalogue displays a non-homogeneous distribution along each segment of the NAF in the Marmara Sea region. The western and eastern segments exhibit 3.8 x $10^{17}$ N.m/yr and 3.6 x $10^{17}$ N.m/yr, respectively, against only 1.8 x $10^{17}$ N.m/yr along the ≃ 100 km of the Central fault segment (Fig. 3). This seismic moment rate difference supports the hypothesis of a seismic gap along the Central Marmara segment, but it does not allow the constraint of its physical limits.

At the scale of all Marmara region, the distribution of the geodetic moment rate shows, on one hand, that dip-slip mechanism is principally located around the Central pull-apart basin reaching a maximum of $\simeq 10^{17}$ N.m/yr, and on the other hand, the strike-slip mechanism along the fault, has an average $\simeq 6 \times 10^{16}$ N.m/yr (Figure 4a and b). Taking into account the existence of geometrical discontinuities along the fault (pull-apart basin, step-over and fault bend) the length of the Central segment may not be limited between Cinarcik basin and the Central basin ($\simeq 70$ km) and may include the Cinarcik fault segment and the section West of the Central basin ($\simeq 130$ km).

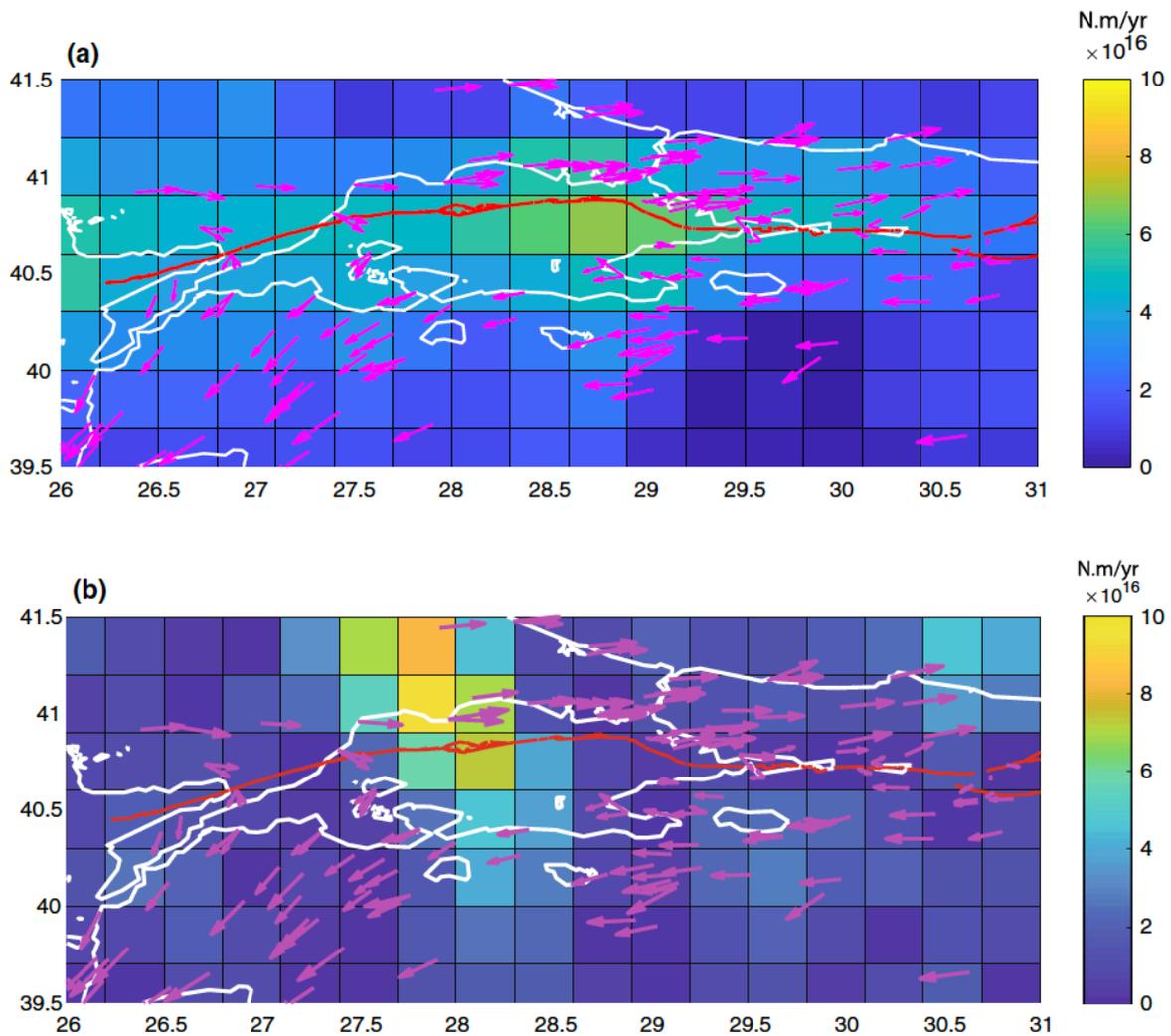

Figure 4. Geodetic moment rate (in N.m/year) obtained from the GPS data in the Marmara Sea (see also Figs. 2 and 6 a and b for scaling). *a* The high geodetic moment rate is linked to the strike-slip component of active deformation along the central fault segment; *b* according to the *dip-slip mechanism, a higher moment rate is located around the Central pull-apart basin. A shift to the North of the geodetic rate may also be partly due to the poor GPS data coverage near the North Anatolian Fault (red line). Pink vectors are inferred from the interpolation of GPS velocities processed to fill the limited GPS data coverage in the Marmara Sea. The interpolation is made according to 0.3° cell distribution.*

The inferred moment rate deficit appears to be localized across the Central basin and includes the Tekirdag basin, Central basin and Central fault segment reaching ≃ 120 km-long with a maximum $\dot{M}_d$ = 6.4 x $10^{17}$ N.m/yr that correlates with an estimated Mw 7.4-7.5 and ≃ 10 mm/yr slip deficit rate (Figs. 5a, b, Table 2). The main difference results in the characteristics of the geodetic rate taking either the strike-slip component (Figs. 4a and 5a) or dip-slip component (Figs. 4b, 5b). However, the strike-slip being the main mechanism along the NAF, the area of the moment deficit helps in the identification of the Central Fault segment that corresponds to the location of the Marmara seismic gap. Considering the moment deficit rate and related average 2.5 to 4 m slip deficit since the May 22, 1766 earthquake, the size of the Central Fault segment needs to amount 130 km-length. These results suggest that the expected earthquake is along the Central Fault Segment of the Marmara Sea and may reach Mw 7.4 within a return period *Ti* of 250-400 years regardless of the contribution of aseismic creeping.

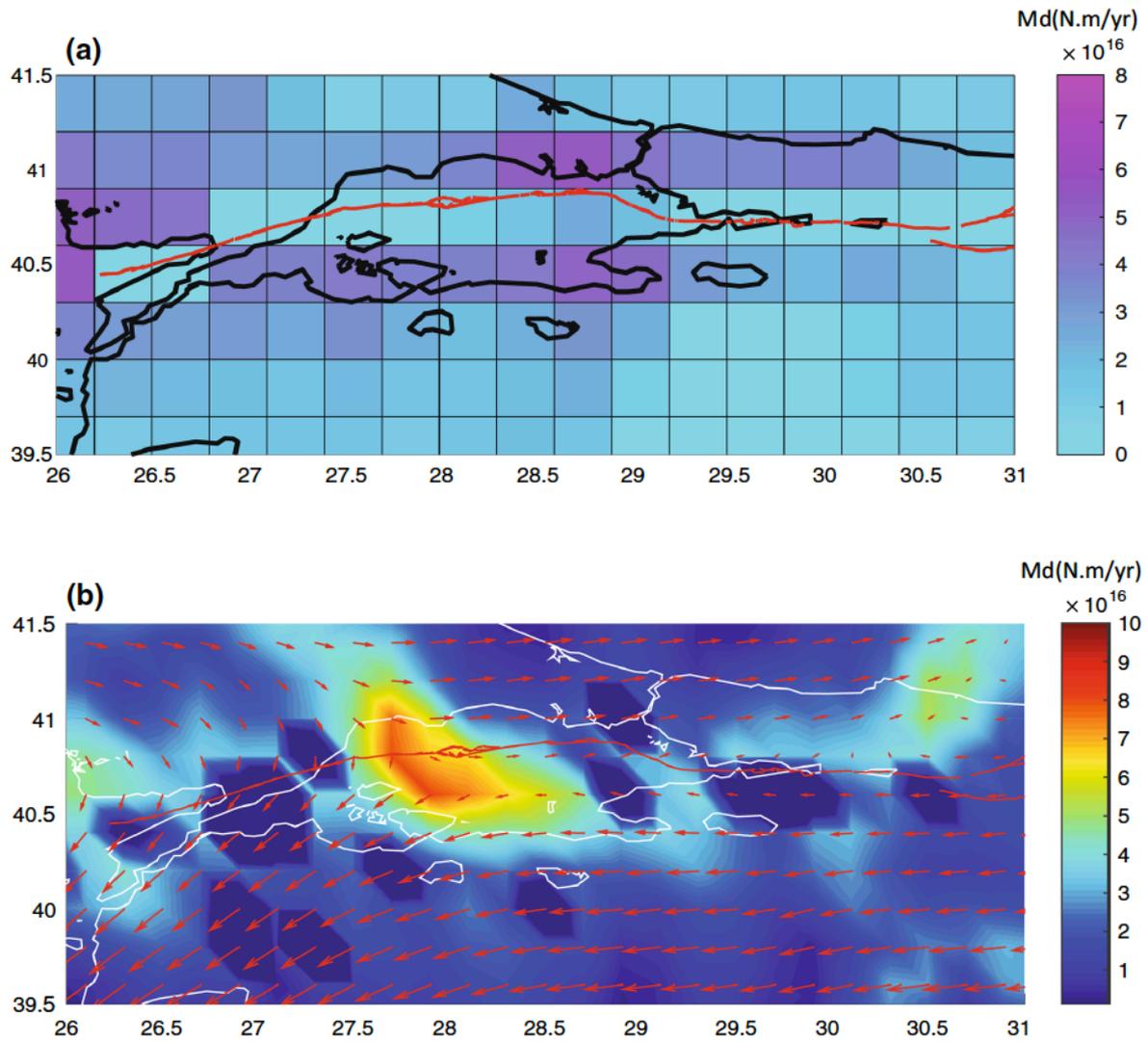

*Figure 5. Moment rate deficit calculated from the difference between seismic and geodetic moment rate along the Marmara fault segments (red line). **a** The high moment rate deficit is associated with the high geodetic rate (see Fig. 4a) and related strike-slip component of active deformation along the central fault segment; **b** The high moment rate is here according to the high geodetic rate and dip-slip mechanism located around the Central pull-apart basin. Red arrows are as in Fig. 4b (see also Fig. 6b for scaling). The moment rate deficit show 6.4 x $10^{17}$ N.m/year and 2.5–4 m slip deficit for Ti = 250 to 400 years return period of large earthquakes along the ~ 130 km-long central Marmara fault segment (see also Table 2)*

# 5 - Discussion and conclusion

The analysis of tectonic, paleoseismological, seismological and geodetic data, allows to constrain the location and size of the seismic gap and related slip deficit along the NAF in the Marmara Sea. The difference between the far field active deformation and the seismic moment rate based on the last 2000 years characterizes the seismic gap in the Central Marmara segment. The paleoseismic and seismicity data indicate that the most recent event in this fault segment took place in May 22, 1766, inferring 2.5 to 4 m seismic slip deficit, $\simeq$ 10 mm/yr seismic rate deficit, correlated with the seismic cycle. The $\simeq$ 130 km-long fault length (15 km-width) suggests an expected earthquake magnitude of Mw $\simeq$ 7.4, comparable to the previous earthquakes along the nearby 1912 and 1999 earthquake rupture segments. Our modelling of the moment rate deficit $\dot{M}_d$ suffers from the lack of GPS data in the Sea of Marmara and related NW regions, which results in relatively high values of $\dot{M}_d$ west and northwest of the Central fault segment. The obtained moment rate deficit show 6.4 x $10^{17}$ N.m/year and 2.5-4 m slip deficit for deficit Ti=250 to 400 years return period of large earthquakes along the ~130 km-long central Marmara fault segment (Fig. 5 and Table 2).

| Ti | $\dot{M}_d$ | Mo (N.m) | Mw | L (km) | W (km) | μ (N/m²) | u (m) | $\dot{u}$ (mm/yr.) |
|---|---|---|---|---|---|---|---|---|
| 251 | 6.41 *$10^{17}$ | 1.60*$10^{20}$ | 7.40 | 130 | 15 | 3.3*$10^{10}$ | 2.5 | 9.96 |
| 300 | 6.41 *$10^{17}$ | 1.92*$10^{20}$ | 7.45 | 130 | 15 | 3.3*$10^{10}$ | 2.99 | 9.96 |
| 350 | 6.41 *$10^{17}$ | 2.24*$10^{20}$ | 7.50 | 130 | 15 | 3.3*$10^{10}$ | 3.49 | 9.96 |
| 400 | 6.41 *$10^{17}$ | 2.56*$10^{20}$ | 7.54 | 130 | 15 | 3.3*$10^{10}$ | 3.98 | 9.96 |

*Table 2: Parameters of the future large earthquake in the Central Marmara fault segment, for different time intervals (Ti), inferred from our modeling of moment rate deficit $\dot{M}_d$.*

The Central Marmara fault segment and related seismic gap, limited to the East and West by the 1999 and 1912 earthquake ruptures, respectively, includes $\simeq$ 70 km-long the Main Marmara fault section, the $\simeq$ 30 km-long Cinarçik fault section, the $\simeq$ 10 km-long Central basin and the ≃ 20 km-long eastern end of the 1912 rupture fault segment. Using the Coulomb failure function modeling, for different rupture scenarios, Pondard et al. (2007) suggest 5 m slip deficit and stress loading along only 70 km-long Main Marmara fault segment. In their analysis of the 1912 coseismic surface rupture and submarine fault scarps, Armijo et al. (2005) and Aksoy et al. (2010) extend the earthquake rupture eastward to the Central basin. However, the comparison between surface ruptures of strike-slip earthquakes shows that rupture lengths are not always dependent on geometrical discontinuities (Wesnousky, 2006), and endpoints of coseismic ruptures may be changing laterally implying the variability of fault segment extremities in the Marmara Sea.

The NAF comprises a ≃creeping section at Ismetpasa on the 1944 rupture segment (Ambraseys and Jackson, 2000). New data of surface deformation characterized by GPS and InSAR time series with inversion modeling identify ≃ 100 km-long creeping section with 8 ± 2 mm/yr and the constraint of a locking depth at 6.5 km (Cakir et al. 2014; Cetin et al., 2014). The occurrence of Mw = 7.3 1944 earthquake was not prevented by the existence of the creeping section. While previous studies suggest an opposition between the occurrence of a large earthquake magnitude and the presence of a creeping section along the Central Marmara fault segment (Ergintav et al., 2014; Schmittbuhl et al., 2016), similarly to the existence of the Ismetpasa creeping section, we consider in our model that both locked-fault rupture and aseismic slip can be coeval. Using a different approach based on GPS slip deficit model, Bulut et al. (2019) suggest comparable results with ~4m slip deficit along the Central Marmara Fault Segment.

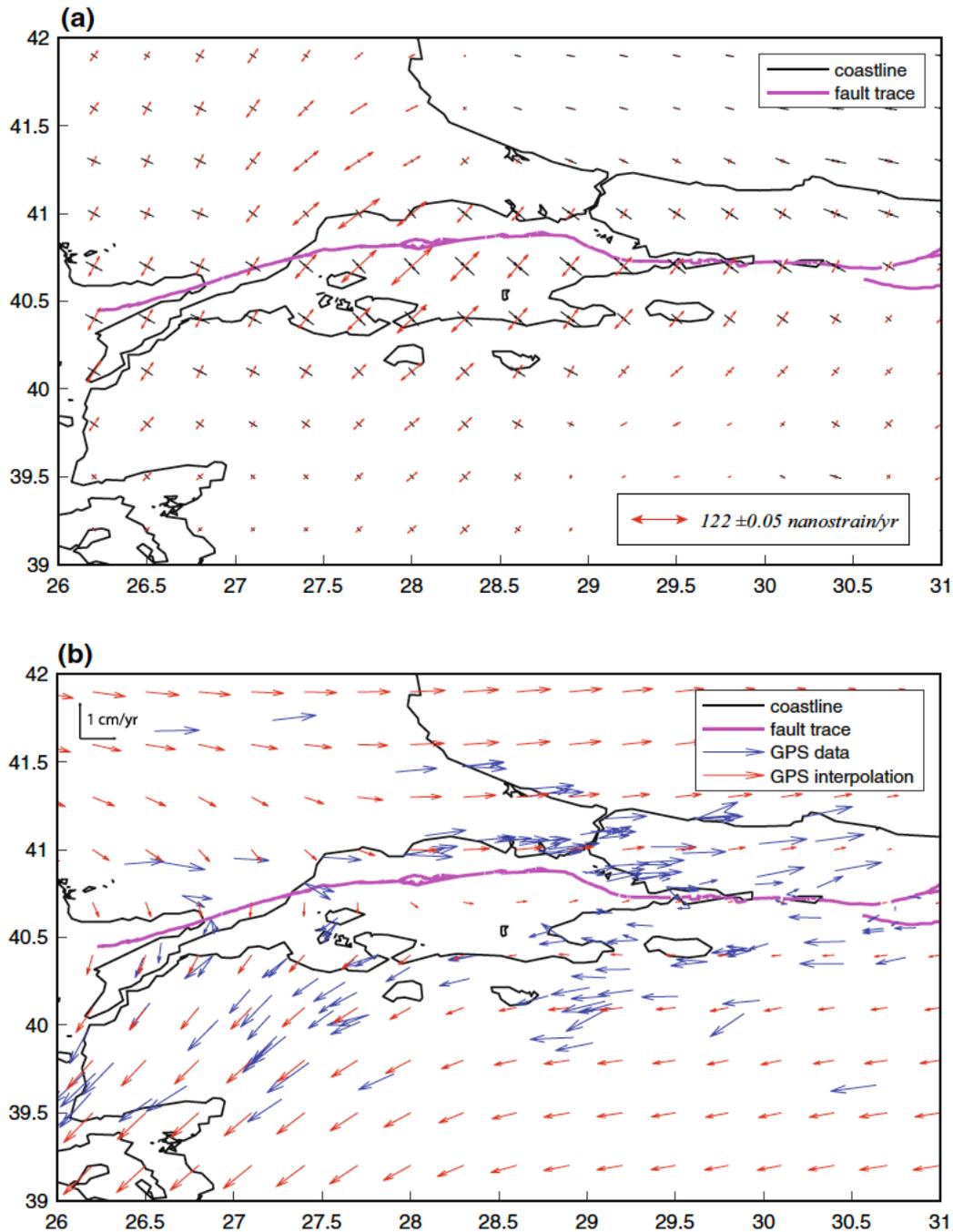

*Figure 6. **a** Surface strain eigenvalues and eigenaxis that amount ~ 122 nanostrain/yr in the Marmara Sea region extracted from the GPS data. Red line is the North Anatolian Fault. **b** GPS data and related interpolation (see text and Supplementary Material for explanation).*

The characterization of the seismic gap and related moment rate deficit is crucial for the hazard and risk assessment of the Istanbul region. The stress loading in the Marmara Sea (Stein et al., 1997; Hubert-Ferrari et al., 2000) and the occurrence of the 1999 Izmit earthquake brought forward the possibility of a future large earthquake in the Sea of Marmara. Our modeling suggest that a large portion of the Central Marmara Sea fault segment is locked, and without excluding a creeping section, the moment rate deficit $\dot{M}_d$ = 6.4 x $10^{17}$ N.m/yr implies the occurrence of a Mw = 7.4 future earthquake with an average ≃ 3.25 m coseismic slip along strike. The paleoseismic results indicating a seismic cycle ranging between 250 and 400 years, and our estimate of the seismic slip deficit being ≃ 10 mm/yr, with the uncertainty in the strain accumulation (122 ±0.05 nanostrain/yr, Fig. 6a), the size of the future earthquake on the 130 km-long Central fault segment ranges between Mw = 7.4 and 7.5.

**Acknowledgments:**

This work is dedicated to Aykut Barka who introduced us to the earthquake activity in the Istanbul and Marmara region. We are thankful to Serdar Akyuz, Erhan Altunel, Ziyadin Cakir, Esra Certin, Aynur Dikbas, Osgur Kozci, Matthieu Ferry, Tom Rockwell, Gulsen Ucarkus, for the discussion on the faulting characteristics of the Marmara region. We are indebted to Semih Ergintav and Hayrullah Karabulut for providing GPS and seismicity data. We thank Aurélie Flamand for her study of deformation rate during her Master degree. This research with paleoseismic data collection was partly supported by the INSU-CNRS and EU-funded project RELIEF EVG1-2002-00069. We are also grateful to two anonymous reviewers who helped improving the manuscript presentation. We also thank the Research Council of Norway through its Centres of Excellence funding scheme, project number 262644. Some figures have been produced using Generic Mapping Tools (Wessel and Smith 1995).

**Author contributions**

MM, RT and MEA did the data analysis. RT did the spatial interpolation of
GPS data. MM and RT did the modeling of slip deficit. MM, RT and MEA

prepared the manuscript.

**Funding**

The INSU-CNRS, EU- project RELIEF EVG1-2002–00069 and the Research

Council of Norway through its Centre of Excellence funding scheme, project

number 262644.

**Availability of data and material**

The instrumental seismicity data obtained from Kandilli Observatory

http://www.koeri.boun.edu.tr/sismo/2/earthquake-catalog/

Earthquake Research Institute and the International Seismological Center,

http://www.isc.ac.uk/iscbulletin/search/catalogue/ .

**Code availability**

On request to the corresponding author.

**Declaration**

Conflict of interest No conflict or competing interest.

# 6. Supplementary Information

Below is the link to the electronic supplementary material.

Supplementary file1 (PDF 113 KB):

https://static-content.springer.com/esm/art%3A10.1007%2Fs42990-021-00053-

w/MediaObjects/42990_2021_53_MOESM1_ESM.pdf